\documentstyle[preprint,aps]{revtex}
\draft
\tightenlines
\begin{document}

\title{Superconducting state properties of a $d$-wave superconductor
with mass anisotropy}

\author{Ji-Hai Xu, Yong Ren, and C. S. Ting}
\address{Texas Center for Superconductivity, University of Houston,
Houston, Texas, 77204}
\maketitle
\begin{abstract}
YBa$_2$Cu$_3$O$_7$ (YBCO) exhibits a large anisotropy between
the $a$ and $b$ axes
in the CuO$_2$ planes because of the presence of CuO chains. In order to
account for such an anisotropy we develop a Ginzburg-Landau (GL) theory
for an anisotropic d-wave superconductor in an external magnetic field,
based
on an anisotropic effective mass approximation within CuO$_2$ planes. The
anisotropic parameter $\lambda=m_x/m_y$, where $m_x$ ($m_y$) is the effective
mass in the $x$ ($y$) direction, is found to have significant physical
consequences: In the bulk case, there exist both the $s$- and $d$-wave
order parameters with the same transition temperature, as long as
$\lambda\ne 1$. The GL equations are also solved both analytically and
numerically for the vortex state, and it is shown that both
the $s$- and $d$-wave components show a two-fold symmetry, in contrast to
the four-fold symmetry around the vortex, as expected for the purely
$d$-wave vortex. With
the deviation of $\lambda$ from unity, the opposite winding between the $s$-
and
$d$-wave components observed in the purely $d$-wave case is gradually
taken over by the same winding number.
The vortex lattice is found to have oblique structure in a wide temperature
range  with the precise shape depending on the anisotropy.

\end{abstract}
\narrowtext

\section{Introduction}
Recently, the order parameter symmetry has become the central issue in
high-$T_c$ superconductivity. Many experiments which probe directly the phase
of the pairing state have provided strong evidence for a sign change of the
order parameter \cite{wollman,tsuei,miller,mathai}, consistent with a
predominantly $d$-wave pairing symmetry. At the same time, there are several
measurements which can not be explained within the simple $d_{x^2-y^2}$
state. For example, finite tunneling current along the $c$ axis of a copper
oxide clearly shows  an $s$-wave character \cite{sun}, because there should be
no Josephson current between a CuO$_2$ plane with a gap of $d_{x^2-y^2}$
symmetry and a conventional $s$-wave superconductor.

It is well known that YBCO is not in the purely tetragonal phase due to
the existence of chains. Indeed, YBCO exhibits a large anisotropy between
the $a$ and $b$ directions in the measurements of the penetration depth
\cite{hardy} and the vortex structure by scanning tunneling microscopy (STM)
\cite{stmv}. It was argued recently \cite{muller} that  these
apparently conflicting experimental results in this material may be explained
by assuming that there exist two order parameters, with different symmetry but
the same transition temperature. Namely, the main gap with a $d$-wave symmetry
would result from the CuO$_2$ planes, and a smaller $s$-wave component would
be due to the CuO chains.

In this work, we will consider a simple model for an anisotropic
$d$-wave superconductor, based on the anisotropic effective mass
approximation within a single CuO$_2$ plane. In this model,
the $a$-$b$ anisotropy of YBCO is
taken into account by a single parameter, namely the electron
mass anisotropy, $\lambda=m_x/m_y$, which can be fit to the measured
penetration depth anisotropy \cite{hardy}.
Then the  Ginzburg-Landau (GL) theory for such an anisotropic
$d$-wave superconductor will be studied. Following the procedure
described in Ref.\cite{ren},
we first derive microscopically the GL equations for
this anisotropic system, and then consider the possible solutions of these
GL equations for both bulk and vortex states.
We will show that the anisotropic parameter $\lambda$ has significant
physical consequences: in the bulk, the stable solution from our theory
is the mixed $s+d$ state, and both the $s$- and $d$-wave order
parameters have the same transition temperature. This $s+d$ state is
just what we want to explain the tunneling data and other apparently
conflicting results observed in YBCO. The GL equations are also solved
both analytically and numerically for the vortex structures. We find
that the anisotropic $d$-wave vortex is very different from the purely
$d$-wave case. Namely, both
the $s$- and $d$-wave components show a two-fold symmetry, in contrast to
the four-fold symmetry around the vortex as expected for the purely
$d$-wave vortex. Specifically, the $d$-wave order parameter exhibits
an elliptic shape and the $s$-wave component shows a shape of
butterfly. With the deviation of $\lambda$ from unity, the opposite
winding between the $s$- and $d$-wave components obtained in the purely
$d$-wave case \cite{ren} is gradually taken over by the same winding number.
The vortex lattice is found to be in oblique in a wide temperature
range  with the precise shape depending on the anisotropic parameter
$\lambda$. Here we wish to point out that the preliminary results of
$\lambda=1$ case for the structures of a single vortex and vortex
lattice were reported recently in a conference on superconductivity
\cite{xu}. Here we concentrate on the results for the $\lambda\ne 1$
case.

In Sec. II, starting from Gorkov's theory of weakly coupled
superconductors \cite{gorkov}, the GL equations for the anisotropic $d$-wave
superconductor are derived.
In Sec.III, we discuss the possible solutions of the GL equations for a
uniform or bulk system. In Sec.IV, we study analytically the
qualitative features of a single vortex
using the GL equations. In Sec.V, we present the numerical result for
single vortex structure. The numerical calculation for a vortex lattice
structure is performed in Sec.VI, and Sec.VII includes conclusion and
discussions.

\section{GL equations}
\renewcommand{\theequation}{2.\arabic{equation}}
\setcounter{equation}{0}

In this section we shall derive the GL equations for the anisotropic
$d$-wave superconductor, following closely the procedure we used for a
purely $d$-wave superconductor \cite{ren}. Here  only the main
steps are presented and the difference between the present work and
the previous one will be ephasized. Our starting point
is the gap equation

\begin{equation}
\Delta^{*}({\bf x, x'})=V({\bf x- x'})T
\sum_{\omega_n}F^{\dag}({\bf x, x'},\omega_n),
\end{equation}
which allows for more general than conventional $s$-wave pairing.
$V({\bf x- x'})$ is the effective two-body interaction of the weak-coupling
theory. Using Gorkov \cite{gorkov} description of a superconductor in the
magnetic field:

\begin{eqnarray}
&&\left[ i\omega_n-{1\over 2m}
(-i{\bf \nabla}+e {\bf A})^2 +\mu\right]
\tilde G({\bf x,x'},\omega_n) +\int d{\bf x}''
\Delta({\bf x,x''}) F^{+}
({\bf x'', x'},\omega_n)=\delta({\bf x-x'}),\\
&&\left[ -i\omega_n-{1\over 2m}
(i{\bf \nabla}+e{\bf A})^2 +\mu\right]
F^{+}({\bf x,x'},\omega_n) +\int d{\bf x}''
\Delta^{*}({\bf x, x''})\tilde G({\bf x'',x'},\omega_n)=0,
\end{eqnarray}
where $\mu$ is the Fermi energy and ${\bf A}$ is the vector potential.
The normal-state Green's function at zero magnetic field
can be written in the form

\begin{eqnarray}
G_0({\bf x},\omega_n)= {1\over (2\pi)^2} \int d{\bf k}
e^{i{\bf k}\cdot{\bf x}}{1\over i\omega_n-\xi_{\bf k}},
\end{eqnarray}
where

\begin{equation}
\xi_{\bf k}=\frac{k_x^2}{2m_x}+\frac{k_y^2}{2m_y} -\mu,
\end{equation}
is the single particle energy measured from the Fermi energy $\mu$. We note
that the case with $m_x=m_y$ in the above equation corresponds
to the isotropic $d$-wave
superconductor. Keeping up to the third order in $\Delta$, defining
the central-mass coordinates ${\bf R}=({\bf x}+{\bf x}')/2$ and
the relative coordinates ${\bf r}={\bf x}-{\bf x}'$, and making
Fourier transform with respect to the relative coordinates, we obtain
\cite{ren}

\begin{eqnarray}
\Delta({\bf R,k})&=&\int \frac{d {\bf k}'}{(2\pi)^2}
V({\bf k'-k})T\sum_{\omega_n}
\frac{1}{\omega_n^2+\xi_{\bf k'}^2}\Delta({\bf R,k'})\nonumber\\
&+&\int {d{\bf k}'\over 2(2\pi)^2}
V({\bf k'}-{\bf k}) T\sum_{\omega_n}
\biggl [
{\xi_{\bf k}'^2-3\omega_n^2\over (\omega_n^2+\xi_{k'}^2)^3}
\left(\frac{k'^2_x}{2m_x^2}\Pi_x^2
+\frac{k'^2_y}{2m_y^2} \Pi_y^2\right)\nonumber\\
&& \;\;\;\;\;\;\;\;\;\;\;\;\;\;\;\;\;\;\;\; \;\;\;\;\;\;\;\;\;\;
\;\;\;\;\;\;\;\;\;\; \;\;\;\;
-{\xi_{\bf k'} \over (\omega_n^2+\xi_{\bf k'}^2)^2}
\left( \frac{\Pi_x^2}{2m_x}+
\frac{\Pi_y^2}{2m_y}\right)
\biggr ]
\Delta ({\bf R, k'})\nonumber\\
&-&\int {d{\bf k}'\over (2\pi)^2}
V({\bf k}-{\bf k}') T\sum_{\omega_n}
{1\over (\omega_n^2+\xi_{\bf k'}^2)^2}
|\Delta({\bf R},{\bf k}')|^2 \Delta({\bf R},{\bf k}'),
\end{eqnarray}
where we have introduced the operator
\begin{equation}
{\bf \Pi}=i\nabla_R-2e{\bf A}_R.
\end{equation}

In order to obtain the generic Ginzburg-Landau equations, which govern the
spatial variation of the order parameters, for an anisotropic
$d$-wave superconductor, we need to specify the form of the interaction. Here
we use the following ansatz for the effective interaction responsible for
the spin-singlet pairing:

\begin{equation}
V({\bf k-k'})=-V_s+V_d(\hat{k}_x^2-\hat{k}_y^2)(\hat{k}'^2_x-\hat{k}'^2_y),
\label{threefour}
\end{equation}
where $\hat{\bf k}={\bf k}/|{\bf k}|$ is the unit vector in the direction of
${\bf k}$. By taking both $V_d$ and $V_s$ positive, the $-V_d$ corresponds
to attractive interaction responsible for $d$-wave pairing, and $V_s$ can be
regarded as an effective on-site repulsive interaction.
The general expression of order parameter that follows Eq.(2.8) is

\begin{equation}
\Delta({\bf R},{\bf k})
=\Delta_s ({\bf R})+\Delta_d({\bf R})(\hat k_x^2-\hat k_y^2).
\end{equation}
Substituting Eqs.(2.8) and (2.9) into Eq.(2.6),
and Comparing both sides of the gap equation for $\hat k$-independent terms and
terms proportional to $(\hat k_x^2-\hat k_y^2)$, we obtain the GL equations in
a form suitable for finding the GL free energy functional:

\begin{eqnarray}
&&\alpha_s\Delta_s-\frac{\lambda -1}{\lambda+1}\Delta_d+2\alpha\gamma_d \mu
      \biggl \{ \left(\frac{\Pi_x^2}{2m_x}+\frac{\Pi_y^2}
      {2m_y}\right)\Delta_s\nonumber\\
&&\;\;\;\;\;\;+ \left[
      \frac{\lambda+2\sqrt{\lambda}-1}{(1+\sqrt{\lambda})^2}
      \frac{\Pi_x^2}{2m_x}-
      \frac{2\sqrt{\lambda}+1-\lambda}{(1+\sqrt{\lambda})^2}
      \frac{\Pi_y^2}{2m_y} \right]\Delta_d\biggr \}\nonumber\\
&&\;\;\;\;\;\;+2\gamma_d\alpha\biggl [ \Delta^*_s\Delta_s^2+
      \frac{\sqrt{\lambda}-1}{\sqrt{\lambda}+1}
       |\Delta_s|^2\Delta_d
      +\frac{\lambda+1}{(1+\sqrt{\lambda})^2}
      \Delta_s^*\Delta_d^2\nonumber\\
&&\;\;\;\;\;\;+\frac{\sqrt{\lambda}-1}{\sqrt{\lambda}+1}\Delta_s^2\Delta_d^*
      +2\frac{\lambda+1}{(1+\sqrt{\lambda})^2}\Delta_s|\Delta_d|^2
      +\frac{\lambda\sqrt{\lambda}-1}{(1+\sqrt{\lambda})^3}
      |\Delta_d|^2\Delta_d\biggr ]=0,
\end{eqnarray}

\begin{eqnarray}
&&\left[ 1-\gamma_d \ln\left(\frac{2e^\gamma\omega_0}{\pi T}\right)
      \frac{2(1+\lambda)}{(1+\sqrt{\lambda})^2}\right]\Delta_d
      -\frac{\lambda -1}{\lambda+1}\Delta_s\nonumber\\
&&\;\;\;\;\;\;+2\alpha\gamma_d \mu
      \biggl \{ \left[
             \frac{1-\sqrt{\lambda}+3\lambda+\lambda\sqrt{\lambda}}
             {(1+\sqrt{\lambda})^3} \frac{\Pi_x^2}{2m_x}
          +\frac{1+3\sqrt{\lambda}-\lambda+\lambda\sqrt{\lambda}}
                           {(1+\sqrt{\lambda})^3}
          \frac{\Pi_y^2}{2m_y}\right] \Delta_d  \nonumber\\
&&\;\;\;\;\;\;+\left[
          \frac{\lambda+2\sqrt{\lambda}-1}{(1+\sqrt{\lambda})^2}
          \frac{\Pi_x^2}{2m_x}-
          \frac{2\sqrt{\lambda}+1-\lambda}{(1+\sqrt{\lambda})^2}
          \frac{\Pi_y^2}{2m_y} \right]\Delta_s
      \biggr \}\nonumber\\
&&\;\;\;\;\;\;+2\gamma_d\alpha
      \biggl [
          \frac{\sqrt{\lambda}-1}{\sqrt{\lambda}+1}
          \Delta_s^2\Delta_s^*+
          2\frac{\lambda +1}{(1+\sqrt{\lambda})^2}
          |\Delta_s|^2\Delta_d+\frac{\lambda\sqrt{\lambda}-1}
          {(1+\sqrt{\lambda})^3}\Delta_s^*\Delta_d^2\nonumber\\
&&\;\;\;\;\;\;+\frac{\lambda+1}{(1+\sqrt{\lambda})^2}
      \Delta_s^2\Delta_d^*+
      +2\frac{\lambda\sqrt{\lambda}-1}{(1+\sqrt{\lambda})^3}
       \Delta_s|\Delta_d|^2
      +\frac{(1+\lambda)(1+\lambda+\sqrt{\lambda})}
       {(1+\sqrt{\lambda})^4}|\Delta_d|^2\Delta_d\biggr ]=0,
\end{eqnarray}
where $\alpha=7\zeta(3)/8(\pi T)^2$, $\gamma$ is the Euler constant,
$\gamma_d=N(0)V_d/2$ is the interaction strength in the purely $d$-wave
channel when $\lambda=1$, and

\begin{equation}
\alpha_s=\frac{(1+\sqrt{\lambda})^2}{1+\lambda}\left( 1+
         \frac{(1+\sqrt{\lambda})^2}{1+\lambda}\frac{V_s}{V_d}\right).
\end{equation}
It is easy to show that if setting $\lambda=1$, Eqs.(2.10) and (2.11)
return back
to the results we obtained for a purely $d$-wave superconductor \cite{ren}.
The corresponding GL free energy is

\begin{eqnarray}
F=&&
\left[ 1-\gamma_d \ln\left(\frac{2e^\gamma\omega_D}{\pi T}\right)
      \frac{2(1+\lambda)}{(1+\sqrt{\lambda})^2}\right]|\Delta_d|^2
      +\alpha_s|\Delta_s|^2-\frac{\lambda-1}{\lambda+1}
      (\Delta_s^*\Delta_d+\Delta_d^*\Delta_s) \nonumber\\
&&+2\gamma_d\alpha\mu
     \biggl [ \frac{|\Pi_x\Delta_s|^2}{2m_x}
      +\frac{|\Pi_y\Delta_s|^2}{2m_y}
      +\frac{1-\sqrt{\lambda}+3\lambda+\lambda\sqrt{\lambda}}
        {(1+\sqrt{\lambda})^3}\frac{|\Pi_x\Delta_d|^2}{2m_x}\nonumber\\
&&\;\;\;\;\;\;\;\;\;\;\;\;\;\;\;\;\;\;\;\;\;\;\;
\;\;\;\;\;\;\;\;\;\;\;\;\;\;\;\;\;\;\;\;\;\;
      +\frac{1+3\sqrt{\lambda}-\lambda+\lambda\sqrt{\lambda}}
        {(1+\sqrt{\lambda})^3}\frac{|\Pi_y\Delta_d|^2}{2m_y}
       \nonumber\\
&&+ \left(
      \frac{\lambda-2\sqrt{\lambda}-1}{(1+\sqrt{\lambda})^2}
      \frac{\Pi_x\Delta_s\Pi_x^*\Delta_d^*}{2m_x}-
      \frac{2\sqrt{\lambda}+1-\lambda}{(1+\sqrt{\lambda})^2}
      \frac{\Pi_y\Delta_s\Pi_y^*\Delta_d^*}{2m_y}+{\rm h.c.}\right)
      \biggr ] \nonumber\\
&&+2\gamma_d\alpha
      \biggl [
       \frac{(\lambda+1)(1+\sqrt{\lambda}+\lambda)}
      {(1+\sqrt{\lambda})^4}|\Delta_d|^4+\frac{1}{2}|\Delta_s|^4
      +\frac{2(1+\lambda)}{(1+\sqrt{\lambda})^2}|\Delta_d|^2|\Delta_s|^2
       \nonumber\\
&&+ \frac{\lambda+1}{2(1+\sqrt{\lambda})^2}
      (\Delta_s^{*2}\Delta_d^2+\Delta_s^2\Delta_d^{*2})
    + \frac{\sqrt{\lambda}-1}{\sqrt{\lambda}+1}|\Delta_s|^2
      (\Delta_s^{*}\Delta_d+\Delta_s\Delta_d^{*})\nonumber\\
&&+\frac{\lambda\sqrt{\lambda}-1}{(1+\sqrt{\lambda})^3}|\Delta_d|^2
      (\Delta_s^{*}\Delta_d+\Delta_s\Delta_d^{*})\biggr ].
\end{eqnarray}

It is interesting to note in the above equation that except for the
mixed gradient terms, which are induced by the magnetic field, there
exist the new terms, such as
\[
\propto (\lambda-1) (\Delta_s^{*}\Delta_d+\Delta_s\Delta_d^{*})
\]
and
\[
\propto (\sqrt{\lambda}-1)\left(|\Delta_s|^2+\frac{1+\sqrt{\lambda}
+\lambda}{(1+\sqrt{\lambda})^2}|\Delta_d|^2\right)
      (\Delta_s^{*}\Delta_d+\Delta_s\Delta_d^{*}).
\]
These new terms come completely from the mass anisotropy. For isotropic
systems with $\lambda=1$, these terms vanish. In the
following section, we will discuss the physical consequences of
these new terms.

\section{bulk solutions}
\renewcommand{\theequation}{3.\arabic{equation}}
\setcounter{equation}{0}

In this section we study the solutions of the GL equations of an
anisotropic $d$-wave superconductor for a bulk or uniform system.
In this case the gradient terms in Eqs.(2.10) and (2.11) are equal to
zero. First let us examine the $T_c$ formula. For $T\rightarrow T_c$,
the coefficients of the linear terms in the GL equations determine
the transition temperature:

\begin{equation}
    	\ln\frac{T_c}{T_{c0}}=\frac{1}{\gamma_d}
           \left\{
                1-\frac{(1+\sqrt{\lambda})^2}{2(1+\lambda)}
                \left[
                     1-\frac{1}{\alpha_s}
                        \left(
                             \frac{\lambda-1}{\lambda+1}\right)^2
           \right]\right\},
\end{equation}
where $T_{c0}$ is the transition temperature  for a purely $d$-wave
superconductor corresponding to $\lambda=1$, which is given by

\begin{equation}
	\ln\frac{2e^{\gamma}\omega_D}{\pi T_{c0}}=\frac{1}{\gamma_d}.
\end{equation}
For a small anisotropy ($\lambda\rightarrow 1$), Eq.(3.1) reduces to

\begin{equation}
\ln\frac{T_c}{T_{c0}}=\frac{1}{4\gamma_d\alpha_s}(\lambda-1)^2>0.
\end{equation}
This result implies that $T_c$ increases as the system deviates from
the isotropy.

In the following discussion, for convenience, we put the GL free
energy into a dimensionless form, which can be done by scaling the
energy by $4\gamma_d(1-T/T_c)^2/3\alpha$, lengths by $\xi_x=\sqrt{
\mu\alpha/2m_x(1-T/T_c)}$, and setting
$\Delta_s=\psi_s\Delta_d^{\infty}(\lambda=1)$,
$\Delta_d=\psi_d\Delta_d^{\infty}(\lambda=1)$, where
$\Delta_d^{\infty}(\lambda=1)=\sqrt{4(1-T/T_c)/3\alpha}$, and
${\bf A}=(2\pi\xi_x/\Phi_0){\cal{ A}}$:

\begin{eqnarray}
F=&-&\frac{2(1+\lambda)}{(1+\sqrt{\lambda})^2}
    \biggl [
            1-\frac{1}{\alpha_s\gamma_d(1-T/T_c)}\left(
            \frac{\lambda-1}{\lambda+1}\right)
    \biggr ] |\psi_d|^2
   +\frac{\alpha_s}{\gamma_d(1-T/T_c)}|\psi_s|^2\nonumber\\
&-&\frac{1}{\gamma_d(1-T/T_c)}\left(\frac{\lambda-1}{\lambda+1}\right)
        \left(\psi_s^*\psi_d+\psi_d^*\psi_s\right)
        +2\left(|\Pi_x\psi_s|^2+\lambda|\Pi_y\psi_s|^2\right)\nonumber\\
&+&2\biggl [
            \frac{1-\sqrt{\lambda}+3\lambda+\lambda\sqrt{\lambda}}
                 {(1+\sqrt{\lambda})^3}|\Pi_x\psi_d|^2+\lambda
            \frac{1+3\sqrt{\lambda}-\lambda+\lambda\sqrt{\lambda}}
                 {(1+\sqrt{\lambda})^3}|\Pi_y\psi_d|^2
    \biggr ]\nonumber\\
&+&2\biggl [
            \frac{\lambda+2\sqrt{\lambda}-1}{(1+\sqrt{\lambda})^2}
            \Pi_x\psi_s\Pi_x^*\psi_d^*-
            \frac{1+2\sqrt{\lambda}-\lambda}{(1+\sqrt{\lambda})^2}
            \Pi_y\psi_s\Pi_y^*\psi_d^*+{\rm h.c.}
    \biggr ]\nonumber\\
&+&\frac{4}{3}\frac{(\lambda+1)(\lambda+\sqrt{\lambda}+1)}{
        (1+\sqrt{\lambda})^4}|\psi_d|^4
      +\frac{4}{3}|\psi_s|^4+\frac{16}{3}\frac{\lambda+1}
        {(1+\sqrt{\lambda})^2}|\psi_d|^2|\psi_s|^2\nonumber\\
&+&\frac{4}{3}\frac{\lambda+1}{(1+\sqrt{\lambda})^2}
        \left(\psi_s^{*2}\psi_d^2+\psi_d^{*2}\psi_s^2\right)
        +\frac{8}{3}\frac{\sqrt{\lambda}-1}{\sqrt{\lambda}+1}
        |\psi_s|^2\left(\psi_s^*\psi_d+\psi_d^*\psi_s\right)\nonumber\\
&+&\frac{8}{3}\frac{\lambda\sqrt{\lambda}-1}{(1+\sqrt{\lambda})^3}
        |\psi_d|^2\left(\psi_s^*\psi_d+\psi_d^*\psi_s\right)
        +\kappa^2(\nabla\times {\cal{A}})^2.
\end{eqnarray}
In the above expression, the magnetic field energy has been included
explicitly with $\kappa$ being the GL parameter. From
$\delta F/\delta\psi_s^*$ and $\delta F/\delta\psi_d^*$
we can obtain the GL equations for $\psi_s$ and $\psi_d$. We now discuss
the bulk solutions. Assuming
$\psi_s=|\psi_s|e^{i\theta_s}$,
$\psi_d=|\psi_d|e^{i\theta_d}$, and $\theta=\theta_s-\theta_d$, we can
determine the value of $\theta$ through
$\partial F/\partial\theta =0$ and $\partial^2 F/\partial\theta^2>0$.
We find that the stable solution is only possible for $\theta=0$, which
means that the bulk system is in the mixed $s+d$ state with a real
combination. For $T\rightarrow T_c$, $\psi_s$ and $\psi_d$ are given by

\begin{equation}
\psi_s=\frac{1}{\alpha_s}\left(\frac{\lambda-1}{\lambda+1}\right)\psi_d,
\end{equation}
\begin{equation}
\psi_d=D^{-1}(\lambda),
\end{equation}
with

\begin{eqnarray}
D(\lambda)=&&\frac{4}{3}\frac{(1+\sqrt{\lambda})^2}{1+\lambda}
          \biggl [
          \frac{(\lambda+1)(\lambda+\sqrt{\lambda}+1)}
                 {(1+\sqrt{\lambda})^4}+
          2\frac{\lambda\sqrt{\lambda}-1}{(1+\sqrt{\lambda})^3}
          \left(\frac{\lambda-1}{\lambda+1}\right)
          \frac{1}{\alpha_s}\nonumber\\
&&+3\frac{\lambda+1}{(1+\sqrt{\lambda})^2}
          \left(\frac{\lambda-1}{\lambda+1}\right)^2
          \frac{1}{\alpha_s^2}
          +\frac{\sqrt{\lambda}-1}{\sqrt{\lambda}+1}
          \left(\frac{\lambda-1}{\lambda+1}\right)^3
          \frac{1}{\alpha_s^3}
          \biggr ].
\end{eqnarray}
It is easy to see that $D(\lambda)\rightarrow 1$ as $\lambda\rightarrow
1$. Namely for an isotropic system with $\lambda\rightarrow 1$,
$\psi_d\rightarrow 1$, and  $\psi_s\rightarrow 0$, which implies that
the purely $d$-wave state is only possible for the isotropic system.
Whenever the system has an anisotropy, the mixed $s+d$ state with
$\psi_d<1$ and $\psi_s\ne 0$ is generated. It is clear from Eq.(3.5)
that these two order parameters have the same $T_c$, which is given by
Eq.(3.1). As argued by many authors \cite{muller}, such a mixed $s+d$
state in the bulk is just what we need to explain the tunneling data and
other apparently conflicting results observed in YBCO.

\section{Vortex solutions}
\renewcommand{\theequation}{4.\arabic{equation}}
\setcounter{equation}{0}

In this section, we determine the single vortex solutions for an
anisotropic $d$-wave superconductor using the GL equations derived in
Sec.II. Previously, we have studied the purely $d$-wave vortex structure
\cite{ren} and found that near the vortex core, there coexist the
$s$-wave and $d$-wave components with the opposite winding numbers.
Far away from the vortex core, the induced $s$-wave component shows
strong four-fold anisotropy and decays as $r^{-2}$. We expect that the
mass-anisotropy will affect the vortex structure.
For simplification, here we study the case when $\epsilon=\lambda-1$ is
a small parameter. In this case the GL equations become

\begin{eqnarray}
       	\frac{\alpha_s}{\gamma_d(1-T/T_c)}\psi_s
      &&-\frac{\epsilon}{2\gamma(1-T/T_c)}\psi_d
        +\frac{8}{3}|\psi_s|^2\psi_s
        +\frac{8}{3}|\psi_d|^2\psi_s
        +\frac{4}{3}\psi_d^2\psi_s^*\nonumber\\
      &&+{2\over 3}\epsilon(|\psi_d|^2\psi_d+\psi_s^2\psi_d^*)
	+{1\over 2}\epsilon|\psi_d|^2\psi_d
  	+2{\bf \Pi}^2\psi_s+(\Pi_x^2-\Pi_y^2)\psi_d=0,
\end{eqnarray}
\begin{eqnarray}
	-\psi_d
      &&+\frac{\epsilon^2}{4\alpha_s\gamma(1-T/T_c)}\psi_d
	-\frac{\epsilon}{2\gamma_d(1-T/T_c)}\psi_s
 	+|\psi_d|^2\psi_d+{8\over 3}|\psi_s|^2\psi_d
        +{4\over 3}\psi_s^2\psi_d^*\nonumber\\
      &&+{2\over 3}\epsilon|\psi_s|^2\psi_s
 	+{1\over 4}\epsilon(\psi_d^2\psi_s^*+|\psi_d|^2\psi_s)
        +{\bf \Pi}^2\psi_d+(\Pi_x^2-\Pi_y^2)\psi_s=0.
\end{eqnarray}

In terms of the cylindrical coordinates, ${\bf R}=(r,\theta)$, we expect
that the $d$-wave component has the form $\psi_d=e^{i\theta}$ in the
region of $1\ll r\ll$ London penetration depth. Also note that, in this
region, the magnetic field effect can be neglected. Then the leading
terms in the equation (4.1) for $\psi_s$ are

\begin{eqnarray}
	\frac{\alpha_s}{\gamma_d(1-T/T_c)}\psi_s
      &&-{1\over 2}\epsilon\left[ \frac{1}{\gamma_d(1-T/T_c)}-1\right]
         e^{i\theta}+{8\over 3}\psi_s\nonumber\\
      &&+{4\over 3}e^{2i\theta}\psi_s^*
 	-\left(\partial_x^2-\partial_y^2\right)e^{i\theta}=0.
\end{eqnarray}
This equation suggests the following solution:

\begin{equation}
	\psi_s=ae^{i\theta}+{1\over r^2}(be^{-i\theta}
	-ce^{3i\theta}),
\end{equation}
where
\begin{equation}
	a=\frac{3\tilde{\epsilon}(3\tilde{\alpha}_s+4)}
               {(3\tilde{\alpha}_s+8)^2-16},
\end{equation}
\begin{equation}
	b={3\over 2}\frac{3\tilde{\alpha}_s+20}
               {(3\tilde{\alpha}_s+8)^2-16},
\end{equation}
\begin{equation}
	c={3\over 2}\frac{9\tilde{\alpha}_s+28}
               {(3\tilde{\alpha}_s+8)^2-16},
\end{equation}
with $\tilde{\alpha}_s=\alpha_s/\gamma_d(1-T/T_c)$ and
$\tilde{\epsilon}=\frac{1}{2}\epsilon[1/\gamma_d(1-T/T_c)-1]$. For
$T\rightarrow T_c$, $\psi_s$ takes the simple expression:

\begin{equation}
	\psi_s=\frac{\epsilon}{2\alpha_s}e^{i\theta}
	      +\frac{\gamma_d(1-T/T_c)}{2\alpha_s}
	       {1\over r^2}\left(e^{-i\theta}-3e^{3i\theta}\right).
\end{equation}
It is very important to note that the first term in the above equation
is independent of both temperature and
the distance from the vortex core. This term comes, in fact,
from the contribution of the bulk. Comparing this term
with the bulk solution given in (3.4), we immediately find that they are
identical for small $\epsilon$ parameter. Our solution (4.4) implies
that far away from the vortex core and the temperature approaches to
$T_c$, the bulk term of the $s$-wave component, with the same winding
with respect to $d$-wave order parameter, becomes dominant.
The magnitude of $\psi_s$ is

\begin{equation}
	|\psi_s|^2=a^2+{1\over r^4}(b^2+c^2)+{2a\over r^2}
	(b-c)\cos 2\theta -{2bc\over r^4}\cos 4\theta.
\end{equation}
This result clearly shows a two-fold symmetry due to the existence of
the $\cos 2\theta$ term.  If setting $\epsilon=0$, the $\cos 2\theta$
term vinishes, $|\psi_s|^2$ recovers the four-fold symmetry, and our result
returns back to that
for a purely $d$-wave  superconductor, as given in Ref.\cite{ren}.

Near the vortex core, to the leading order, our GL equations become

\begin{equation}
-\psi_d-\nabla^2\psi_d=0,
\end{equation}
\begin{equation}
\frac{\alpha_s}{\gamma_d(1-T/T_c)}\psi_s
-\frac{\epsilon}{2\gamma_d(1-T/T_c)}\psi_d
-2\nabla^2\psi_s-\left(\partial^2_x-\partial^2_y\right)\psi_d=0.
\end{equation}

{}From (4.10) we have
\begin{equation}
	\psi_d=c_0\left(r-{1\over 8}r^3\right)e^{i\theta},
\end{equation}
where $c_0$ is a constant. Putting the above equation into (4.11), we
obtain

\begin{equation}
\psi_s=	\frac{\epsilon}{2\alpha_s}c_0re^{i\theta}-
        \frac{\gamma_d(1-T/T_c)}{2\alpha_s}c_0re^{-i\theta}.
\end{equation}
Thus, the leading order terms of the order parameters near the vortex
core are

\begin{eqnarray}
&&	\psi_d=c_0re^{i\theta}\\
&&	\psi_s=\frac{c_0r}{2\alpha_s}\left[\epsilon e^{i\theta}
        -\gamma_d(1-T/T_c)e^{-i\theta}\right].
\end{eqnarray}

For isotropic systems with $\epsilon=0$, the above results reduce to
those for a purely $d$-wave superconductor. Namely, the $s$-wave
component, with the opposite winding relative to the $d$-wave order
parameter, is induced near the vortex core \cite{ren}. However, the
anisotropy alters such a picture: As $T\rightarrow T_c$, the opposite
winding of the $s$-wave component is gradually taken over by the same
winding term. Also $e^{i\theta}$ and $e^{-i\theta}$ terms in Eq.(4.15)
combine to give a two-fold symmetry around the vortex core.

\section{Numerical results of single vortex}
\renewcommand{\theequation}{5.\arabic{equation}}
\setcounter{equation}{0}

In last section, we have discussed analytically the asymptotic behavior
of the single vortex for an anisotropic $d$-wave superconductor. But the
precise shape of the vortex structure is still not clear and it has to
rely on the numerical calculation. Here, we perform a numerical study of
the discretized GL free energy (3.4) using numerical relaxation approach
\cite{alder,wang}. In order to minimize the GL free energy functional
in the presence of magnetic field, we use the
constraint of fixing the average magnetic induction ${\bf B}$ by specifying
the total flux $\Phi$ in the unit cell, and impose the so-called
``magnetic periodic boundary conditions" \cite{wang,doria}.

To perform the numerical relaxation calculation, we need to discretize the
GL free energy (3.4) first. With the use of the forward difference
approximation for the derivatives and taking into account the gauge
invariance, we can write the free energy (3.4) in the discrete form:

\begin{equation}
F=F_0+F_{\rm kin}+F_{\rm field},
\end{equation}
where
\begin{eqnarray}
F_0&&=\frac{1}{N_xN_y}\sum_{ij}\frac{2(1+\lambda)}{(1+\sqrt{\lambda})^2}
    \biggl [
            1-\frac{1}{\alpha_s\gamma_d(1-T/T_c)}\left(
            \frac{\lambda-1}{\lambda+1}\right)
    \biggr ] |\psi_d(i,j)|^2\nonumber\\
&&+\frac{\alpha_s}{\gamma_d(1-T/T_c)}|\psi_s(i,j)|^2
       -\frac{1}{\gamma_d(1-T/T_c)}\left(\frac{\lambda-1}{\lambda+1}\right)
        \left[\psi_s^*(i,j)\psi_d(i,j)+\psi_d^*(i,j)\psi_s(i,j)\right]
         \nonumber\\
&&+\frac{4}{3}\frac{(\lambda+1)(\lambda+\sqrt{\lambda}+1)}{
        (1+\sqrt{\lambda})^4}|\psi_d(i,j)|^4
      +\frac{4}{3}|\psi_s(i,j)|^4+\frac{16}{3}\frac{\lambda+1}
        {(1+\sqrt{\lambda})^2}|\psi_d(i,j)|^2|\psi_s(i,j)|^2\nonumber\\
&&+\frac{4}{3}\frac{\lambda+1}{(1+\sqrt{\lambda})^2}
        \left[\psi_s^{*2}(i,j)\psi_d^2(i,j)+\psi_d^{*2}(i,j)
         \psi_s^2(i,j)\right]\nonumber\\
&&+\frac{8}{3}\frac{\sqrt{\lambda}-1}{\sqrt{\lambda}+1}
        |\psi_s(i,j)|^2\left[\psi_s^*(i,j)\psi_d(i,j)
        +\psi_d^*(i,j)\psi_s(i,j)\right]\nonumber\\
&&+\frac{8}{3}\frac{\lambda\sqrt{\lambda}-1}{(1+\sqrt{\lambda})^3}
        |\psi_d(i,j)|^2\left[\psi_s^*(i,j)\psi_d(i,j)
        +\psi_d^*(i,j)\psi_s(i,j)\right],
\end{eqnarray}
\begin{eqnarray}
F_{\rm kin}=\frac{2}{N_xN_y}\sum_{ij}
    \biggl \{
      \biggl [&&
	|\psi_s(i+1,j)-\psi_s(i,j)e^{ia_xA_x(i,j)}|^2/a_x^2\nonumber\\
&& +\lambda
	|\psi_s(i,j+1)-\psi_s(i,j)e^{ia_yA_y(i,j)}|^2/a_y^2\nonumber\\
&& +\frac{1-\sqrt{\lambda}+3\lambda+\lambda\sqrt{\lambda}}
	        {(1+\sqrt{\lambda})^3}
	|\psi_d(i+1,j)-\psi_d(i,j)e^{ia_xA_x(i,j)}|^2/a_x^2\nonumber\\
&& +\lambda\frac{1+3\sqrt{\lambda}-\lambda+\lambda\sqrt{\lambda}}
	        {(1+\sqrt{\lambda})^3}
	|\psi_d(i,j+1)-\psi_d(i,j)e^{ia_yA_y(i,j)}|^2/a_y^2
      \biggr ]\nonumber\\
&& +\biggl [
   	\frac{\lambda+2\sqrt{\lambda}-1}{(1+\sqrt{\lambda})^2}
	\left(\psi_s(i+1,j)-\psi_s(i,j)e^{ia_xA_x(i,j)}\right)\nonumber\\
&&\;\;\;\;\;\;\;\;\;\;\;\;\;\;\;\;\;\;\;\;\;\;\;\;\times
 	\left(\psi_d^*(i+1,j)-\psi_d^*(i,j)e^{-ia_xA_x(i,j)}\right)
        /a_x^2\nonumber\\
&&-\frac{2\sqrt{\lambda}+1-\lambda}{(1+\sqrt{\lambda})^2}
	\left(\psi_s(i,j+1)-\psi_s(i,j)e^{ia_yA_y(i,j)}\right)\nonumber\\
&&\;\;\;\;\;\;\;\;\;\;\;\;\;\;\;\;\;\;\;\;\;\;\;\;\times
	\left(\psi_d^*(i,j+1)-\psi_d^*(i,j)e^{-ia_yA_y(i,j)}\right)
        /a_y^2+{\rm h.c.}
     \biggr ]
   \biggr \},
\end{eqnarray}
\begin{equation}
F_{\rm field}=\frac{\kappa^2}{N_xN_y}\sum_{ij}
       \biggl \{
	[A_y(i+1,j)-A_y(i,j)]/a_x-[A_x(i,j+1)-A_x(i,j)]/a_y
	\biggr \}^2,
\end{equation}
where $N_x$ ($N_y$) is the number of lattice points in the $x$ ($y$) direction.
On each lattice point $(i,j)$, the order parameters $\psi_s$ and $\psi_d$
have the values $\psi_s(i,j)$ and $\psi_d(i,j)$, and each point is associated
with horizontal and vertical bonds. $a_x$ and $a_y$ are the lattice constants
and $A_x(i,j)$ and $A_y(i,j)$ are the vector potential components on bonds
$[(i,j)\rightarrow (i+1,j)]$ and
$[(i,j)\rightarrow (i,j+1)]$, respectively.
It is easy to show that in these lattice notations, the above expressions
are invariant with respect to gauge transformation:
\begin{eqnarray}
&&\psi_{s,d}(i,j)\rightarrow \psi_{s,d}(i,j)e^{i\chi(i,j)},\nonumber\\
&&A_x(i,j)\rightarrow A_x(i,j)+[\chi(i+1,j)-\chi(i,j)]/a_x,\nonumber\\
&&A_y(i,j)\rightarrow A_y(i,j)+[\chi(i,j+1)-\chi(i,j)]/a_y,\nonumber
\end{eqnarray}
where $\chi(i,j)$ is the arbitrary phase of the order parameters at
site $(i,j)$.
Accordingly, the free energy and other physical quantities are also gauge
invariant. To obtain simple boundary conditions we can choose a gauge such
that $A_x$ is independent of $x$. In this case, our boundary conditions are
\cite{doria,wang}

\begin{eqnarray}
&&A_x(0)=A_x(L_y),\\
&&A_y(L_x,y)-A_y(0,y)=\Phi/L_y,\\
&&A_y(x,Ly)=A_y(x,0),\\
&&\psi_{s,d}(x,L_y)=\psi_{s,d}(x,0)e^{i\Phi/2},\\
&&\psi_{s,d}(L_x,y)=\psi_{s,d}(0,y)e^{iy\Phi/L_y},
\end{eqnarray}
where $L_x=N_xa_x$ and $L_y=N_xa_y$. In order to study the single vortex
structure, we can choose one quantum of flux (i.e., $\Phi=2\pi$) in a square
unit cell with $N\times N$ lattice points. With Eqs.(5.1)-(5.4) and the
above boundary conditions, we can now realize the relaxation procedure.
Choosing $\psi_s$, $\psi_s^*$,
$\psi_d$, $\psi_d^*$, ${\cal{A}}_x$, and  ${\cal{A}}_y$ as independent
variables, we can write down the relaxation iteration equations:

\begin{eqnarray}
&&\psi_s^{(n+1)}(i,j)=\psi_s^{(n)}(i,j)
-\epsilon_1\frac{\partial F}{\partial
\psi_s^*(i,j)}\bigg|^{(n)},\\
&&\psi_d^{(n+1)}(i,j)=\psi_d^{(n)}(i,j)
-\epsilon_2\frac{\partial F}{\partial
\psi_d^*(i,j)}\bigg|^{(n)},\\
&&{\cal{A}}_x^{(n+1)}(i,j)={\cal{A}}_x^{(n)}(i,j)
-\epsilon_3\frac{\partial F}{\partial
{\cal{A}}_x(i,j)}\bigg|^{(n)},\\
&&{\cal{A}}_y^{(n+1)}(i,j)={\cal{A}}_y^{(n)}(i,j)
-\epsilon_4\frac{\partial F}{\partial
{\cal{A}}_y(i,j)}\bigg|^{(n)},
\end{eqnarray}
where $\epsilon$'s are all positive numbers to be adjusted to optimize the
convergence rate and $n$ is an integer labeling the generations of iteration.
It has been shown mathematically that $F$ will monotonically decrease to its
optimum state as $n$ increases as long as we choose a proper initial
state \cite{alder}.

In our numerical calculation, the parameters chosen are $N_x=N_y=101$,
$a_x=a_y=0.2\xi_x$, $\kappa=2$, $T=0.5T_c$, and $V_s=0$.
With these parameters, the external magnetic field corresponds
approximately to the thermal critical field $H_c$. The use of the
different parameters does not alter the qualitative physics. Let us first
show the results for an isotropic $d$-wave vortex structure with $\lambda=1$.
Figs.1 and 2 are  typical surface plots for the distribution of
the $d$-wave order parameter and local magnetic field around the vortex,
respectively, which look like the conventional $s$-wave vortex. But if looking
at them closely, we find the difference from the conventional $s$-wave vortex.
Fig.3 is the contour plot of the $d$-wave order parameter. It is clear that
$|\psi_d|$ exhibits a four-fold symmetry. The local magnetic field also shows
a similar four-fold anisotropy (not shown in the contour plot).

The most interesting feature of a single vortex is that a small
$s$-wave component is induced around the core, as shown in Fig.4
(A) (surface plot) and (B) (contour plot). One can clearly see that the
distribution of $|\psi_s|$ exhibits the profile in the shape of a four-leafed
clover, which is in agreement with our analytical result \cite{ren}.
We believe that the presence of this four-fold symmetric $s$-wave component
is the reason to cause four-fold symmetry of
the $d$-wave order parameter $|\psi_d|$ (see Fig.3) and the local magnetic
field $h$ around the vortex.

We now discuss the anisotropic case with $\lambda\ne 1$. With the deviation of
$\lambda$ from unity, we find that both $|\psi_d|$ and $h$ begin to show
a two-fold symmetry, in contrast to the four-fold symmetry as expected for
a purely $d$-wave vortex. Specifically, both of them exhibit an elliptic
shape. Fig.5 is the contour plot for $|\psi_d|$ with $\lambda=2$, which
corresponds approximately to the measured panetration depth anisotropy
\cite{hardy}.

The $s$-wave component is much more sensitive to the anisotropic parameter
$\lambda$. Fig.6 (surface plot) and Fig.7 (contour plot) show how $|\psi_s|$
changes with the increase of $\lambda$. It is apparent that
$|\psi_s|$ exhibits the two-fold symmetry, and
its shape changes from the four-leafed clover in the purely $d$-wave case
($\lambda=1$) to a butterfly as $\lambda$ increases.
These results agree well with our analytical calculation [see Eqs.(4.8)
and (4.15)].

Recently, the vortex structure of YBCO has been directly observed using
STM imaging technique \cite{stmv}. An elongated shape of the vortex was
realized. The ratio of the axes in the apparent elliptic shape is
about 1.5. Furthermore, this elongation was found to be independent of
the scanning direction of the STM tip. We believe that this elongation
directly reflects the $a$-$b$ anisotropy. This observed vortex shape can
be qualitatively account for by the present GL theory for an anisotropic
$d$-wave superconductor (see Fig.3).

\section{Numerical results of vortex lattice}
\renewcommand{\theequation}{6.\arabic{equation}}
\setcounter{equation}{0}

Recent observation of an oblique vortex lattice structure in YBCO has
been reported with the angle between the primitive axes $\beta\sim 73^0$
by the neutron scattering \cite{keimer} and $\beta\sim 77^0$ by STM
measurements \cite{stmv}.
The rich and complicated structure of a single vortex in an anisotropic
$d$-wave superconductor obtained in the previous sections will be
expected to form a different vortex lattice than the conventional
$s$-wave superconductor, and may provide an explanation to the oblique
vortex lattice structure observed in YBCO. To check this,
the vortex lattice structure is going to be studied using numerical
relaxation method. The vortex
lattice structure is still described by the discrete GL free energy
functional given in Eqs.(5.1)-(5.4). We chose a rectangular unit cell
with two vortices \cite{doria}. The periodic boundary conditions, very
similar to Eqs.(5.5)-(5.9) except for $\Phi=4\pi$ in the present case,
are used in our calculations. The ratio of $a_y/a_x$ controls the
shape of the vortex lattice structure. For example, $a_y/a_x=1$
corresponds to the square, while $a_y/a_x=\sqrt{3}$ corresponds to
triangular lattice. We have calculated the dependence of the free
energy on the ratio of $a_y/a_x$ using the same set of parameters
as for the single vortex.

Let us first present the results for the isotropic systems.
Fig.8 displays the free energy as a function of $a_y/a_x$ for the
isotropic $d$-wave superconductor ($\lambda=1$). It is
evident from the figure that the minimum of the free energy
is at the position with $a_y/a_x\sim 1.3$,
signaling that an oblique vortex lattice with the angle $\beta\sim
75^0$ between the primitive axes is stable. In this case, the
superconducting state in the bulk or uniform system is purely
$d$-wave. The $s$-wave component near the vortex core is
induced due completely to the mixed gradient terms in the GL free
energy. To correlate between the single vortex structure and the vortex
lattice, we have performed the calculations for different temperatures.
We find that in a wide temperature range below $T_c$, the presence of
a sizable induced $s$-wave component causes four-fold symmetry of the
$d$-wave order parameter and the local magnetic field around the vortex.
Such an anisotropic single vortex tend to form an oblique lattice
structure. Fig.9 shows the oblique vortex lattice formed at $T/T_c=0.5$
by the $d$-wave order parameter (A) and the $s$-wave component (B) for
the isotropic case ($\lambda=1$). The local magnetic field distribution
is very similar to the $d$-wave order parameter (not shown in the
figure).

However, when $T\rightarrow T_c$,
the induced $s$-wave component is strongly suppressed, which
can be also seen from our analytical results given in Eqs.(4.8)
and (4.13). In this case, the induced $s$-wave component is
too small to affect the distribution of the $d$-wave order parameter
and the local magnetic field. Consequently, both $\psi_d$ and $h$ have
isotropic distribution around the vortex core, similar to the
conventional $s$-wave vortex. These
isotropic isolated vortices prefer to have a triangular vortex lattice,
identical to the vortex lattice in an $s$-wave superconductor
\cite{kleiner}, as expected. This result is confirmed by our numerical
calculation for the free energy that the minimum of the free energy
moves to $a_y/a_x=\sqrt{3}$, i.e., the triangular lattice is stablized.

We now turn to the anisotropic case. Recent STM measurements in YBCO
reveaved an oblique vortex lattice with $\beta\sim 77^0$ \cite{stmv}.
Moreover, the elongated vortex cores with the ratio of principle axes
about 1.5 were also found by this technique. As noted by the authors
of Ref.\cite{stmv}, if this
elongation reflects the intrinsic $a$-$b$ anisotropy in a
conventional $s$-wave superconductor, such an anisotropy
would lead to a distorted vortex lattice with an angle $\beta=82^0$
inconsistent
with the observed value. Thus it seems that the $a$-$b$ anisotropy alone
can not explain the observed vortex lattice structure and additional
effects, such as the order parameter symmetry, must be involved in order
to account for the experimental data. We should note also that although
the vortex lattice in a purely $d$-wave superconductor has an angle
\cite{xu,kallin} very close to the observed value,
the four-fold symmetric vortex cores
obtained in this case are inconsistent with the observed elliptic
vortex cores. In this regard, it is interesting to study the vortex
lattice structure in an anisotropic $d$-wave superconductor which
contains both the $a$-$b$ anisotropy as well as the $d$-wave order
parameter symmetry. Fig.10 shows the free energy as a function of
$a_y/a_x$ for  $\lambda=2$, which corresponds approximately to the
experimental data on the penetration depth \cite{hardy} and the
coherence length \cite{stmv}. It is clear that the minimum of the free
energy locates at $a_y/a_x\sim 1.3$, almost the same position as for
the isotropic $d$-wave superconductor.

We should mention that in an anisotropic $d$-wave superconductor, the
the $a$-$b$ anisotropy and the order parameter symmetry play very
different roles in determining the vortex structures. The vortex tends
to have a two-fold symmetry due to the $a$-$b$ anisotropy, while it
tends to have a four-fold symmetry due to the $d$-wave pairing state.
For small $\lambda$, the $d$-wave order parameter symmetry is important,
and the anisotropy becomes dominant for large anisotropy, as long as
the effect on the vortex structure is concerned. We would like to
point out here that the similar oblique vortex lattice obtained in both
isotropic ($\lambda=1$) and anisotropic ($\lambda=2$) $d$-wave
superconductors is only a coincidence. In general, the real single
vortex and the vortex lattice structures are determined by the
competition between the anisotropy and the $d$-wave order parameter
symmetry.

Fig.11 shows the vortex lattice formed by the $d$-wave (A) and $s$-wave
(B) order  parameters for  an anisotropic $d$-wave superconductor with
$\lambda=2$. The local magnetic field distribution (not shown in the
figure) is very similar to the $d$-wave component. It is seen that the
vortex lattice is oblique with $\beta\sim 75^0$ and the vortex cores are
elliptic. These results are in perfect agreement with the STM
measurements on the vortex lattice structure in YBCO \cite{stmv}.

\section{Conclusions}

We have established microscopically a GL theory for an anisotropic
superconductor with $d_{x^2-y^2}$-wave pairing symmetry within the
anisotropic effective mass approximation.
The experimental basis of our model is recent measurements which
revealed a large anisotropy of the penetration depth \cite{hardy}
and the coherence length \cite{stmv} between $a$ and $b$ directions
in YBCO due to the existence of the CuO chains. In our model, a single
anisotropic parameter $\lambda$ is introduced which gives a measure
of the difference in the two effective masses $m_x$ and $m_y$ and this
parameter can be fitted to the measured penetration depth and the
coherence length data. The GL equations obtained in the present work
should be useful to study the various properties of YBCO.

We have considered the solution of the GL equations for a uniform or
bulk system and found that the stable solution is the mixed $s+d$ state,
and both the $s$- and $d$-wave order parameters have the same transition
temperature. Such an $s+d$ state is just what we need to explain the
tunneling data \cite{sun}
 and the other apparently conflicting experimental data
observed in YBCO.

We also solved the GL equations both analytically and numerically for the
vortex structures. For the single vortex we find
that the anisotropic $d$-wave vortex is very different from the purely
$d$-wave case. Namely, both
the $s$- and $d$-wave components show a two-fold symmetry, in contrast to
the four-fold symmetry around the vortex as expected for the purely
$d$-wave vortex. Specifically, the $d$-wave order parameter exhibits
an elliptic shape and the $s$-wave component shows a shape of
butterfly. With the deviation of $\lambda$ from unity, the opposite
winding between the $s$- and $d$-wave components observed in the purely
$d$-wave case is gradually taken over by the same winding number.

The vortex lattice is in general oblique for both the purely $d$-wave
and the anisotropic $d$-wave superconductors. Although the angle between
the primitive vectors of the vortex lattice in a purely $d$-wave
superconductor is comparable to the observed value, the shape of the
vortex cores are very different from the experiments. On the other hand,
for sn anisotropic $d$-wave superconductor, the shape of the vortex
lattice is determined by the competition between the anisotropy and the
$d$-wave order parameter symmetry. By using the anisotropic parameter
obtained from the experimental data on the penetration depth and the
coherence length, we were able to find a vortex lattice which agree
well with experiments not only in angle between the primitive axes but
also the elliptic shape of the vortex cores.
\vspace{-0.1in}
\acknowledgements
\vspace{-0.1in}
This research was supported by
the State of Texas through the Texas Center for Superconductivity
at the University of Houston, the Texas Advanced Research
Program and the Robert A. Welch Foundation.

\begin{figure}
Fig.1\ Surface plot for the distribution of $|\psi_d|$ around a single
vortex in the isotropic $d$-wave superconductor.The parameters used are
given in the text.
\end{figure}

\begin{figure}
Fig.2\ Surface plot for the distribution of the local magnetic field
$h$ around a single vortex in the isotropic $d$-wave superconductor.
\end{figure}

\begin{figure}
Fig.3\ Contour plot for the distribution of $|\psi_d|$ around a single
vortex in the isotropic $d$-wave superconductor.
\end{figure}

\begin{figure}
Fig.4 \ Distribution of $|\psi_s|$ around a single
vortex in the isotropic $d$-wave superconductor. (A) Surface plot, and
(B) contour plot.
\end{figure}

\begin{figure}
Fig.5\ Contour plot for the distribution of $|\psi_d|$ around a single
vortex in the anisotropic $d$-wave superconductor with $\lambda=2$.
\end{figure}

\begin{figure}
Fig.6\ Surface plot for the distribution of $|\psi_s|$ around a single
vortex in the anisotropic $d$-wave superconductor with different
anisotropic parameters: (A) $\lambda=1.05$, (B) $\lambda=1.2$, and (C)
$\lambda=2$.
\end{figure}

\begin{figure}
Fig.7\ Contour plot for the distribution of $|\psi_s|$ around a single
vortex in the anisotropic $d$-wave superconductor with different
anisotropic parameters: (A) $\lambda=1.05$, (B) $\lambda=1.2$, and (C)
$\lambda=2$.
\end{figure}

\begin{figure}
Fig.8\ Free energy as a function of the ratio of $a_y/a_x$ for vortex
lattice in an isotropic $d$-wave superconductor.
\end{figure}

\begin{figure}
Fig.9\ Contour plots of the $d$- (A) and $s$-wave (B) order parameters
for the stable vortex lattice structure in an isotropic $d$-wave
superconductor with $a_y/a_x=1.3$, corresponding to an oblique lattice
with $\beta=75^0$.
\end{figure}

\begin{figure}
Fig.10\ Free energy as a function of the ratio of $a_y/a_x$ for vortex
lattice in an anisotropic $d$-wave superconductor with $\lambda=2$.
\end{figure}

\begin{figure}
Fig.11\ Contour plots of the $d$- (A) and $s$-wave (B) order parameters
for the stable vortex lattice structure in an isotropic $d$-wave
superconductor ($\lambda=2$) with $a_y/a_x=1.3$, corresponding to an
oblique lattice with $\beta=75^0$.
\end{figure}
\end{document}